\documentstyle[preprint,aps]{revtex}
\begin{document}
\draft
\preprint{hep-th/9510076, SNUTP/95-092}
\title{ Quantum Mechanics of
Integrable Spins on Coadjoint Orbits }
\author  {Sang-Ok Hahn\cite{hahn} and Phillial Oh\cite{poh}}
\address{Department of Physics\\
Sung Kyun Kwan University\\
Suwon 440-746,  KOREA}
\author {Myung-Ho Kim\cite{kim}}
\address{Department of Mathematics\\
Sung Kyun Kwan University\\
Suwon 440-746,  KOREA}

\maketitle

\begin{abstract}
We investigate  classical integrable spins defined on
the reduced phase spaces of coadjoint orbits of $G=
SU(N)$ and study quantum mechanics of them.
 After discussions on a complete set of commuting
functions on each orbit and construction of integrable spin
models  on the flag manifolds,
we quantize a concrete example of integrable spins on SU(3)
flag manifold in the coherent state quantization scheme
and solve explicitly the time-dependent Schr\"odinger equation.
\end{abstract}


\newpage
\def\theequation{\arabic{section}.\arabic{equation}}
\section{Introduction}
\setcounter{equation}{0}
The classical formulation  of non-relativistic spin degrees of
freedom started with the work in Ref. \cite{schu}  by describing
them on the phase space of  $S^2$.
Then, in an independent development of classical isospin
particles, the equations of motion for isospin particles
in the presence of external gauge field were written
down \cite{wong} and the Lagrangian \cite{bala}
and Hamiltonian \cite{ster,kimoh} formulations  for the classical
isospin were accomplished.
Recently, the path integral quantization for the spin was
proposed  \cite{niel,alek2,john,macm}  and it was
extended to formulations for the coherent state path integral
\cite{ston} of generalized $SU(N)$ spin \cite{oh,oh1}.
Especially in Ref. \cite{oh1}, the coherent state path integral
of an integrable spin model  on a flag manifold was performed
and an exact result was obtained for a special case.

In this paper, we further develop  the classical spin theory
on the  coadjoint orbits \cite{kiri} of $G=SU(N)$ group each of which is
a reduced phase space of generalized spin degree of freedom
and study the quantum mechanics of integrable systems on them.
We use the coherent state quantization method \cite{klau}
as our main tool because coupled with geometric quantization method,
it provide a natural way to
quantize the system in terms of the time-dependent Schr\"odinger equation
which can be expressed as a single component wave equation.
This approach is to be contrasted to the
quantization scheme with unitary representation of
coadjoint orbits of compact groups \cite{kiri} which deals, in general,
multi-component equation.

Let us start by giving a brief summary of symplectic structure
on the coadjoint orbits. The configuration space for
the  generalized spin degrees of freedom is
 a Lie group  $G=SU(N)$. Consider the cotangent bundle
$T^*G\cong G\times {\cal G}^*$, where ${\cal G}^*$
is the dual of the Lie algebra ${\cal G}$ of the
group $G$ \cite{arno}.
There is a natural symplectic group action on $T^*G$ via
\begin{eqnarray}
G\times (G\times {\cal G}^*)&\longrightarrow&
G\times {\cal G}^*\nonumber\\
(g, (h,a))&\mapsto& (gh,a).
\end{eqnarray}
Let us define the moment map
 $\rho: T^*G\rightarrow {\cal G}^*$ via
\begin{equation}
<X, \rho(m)>=m\left(\frac{d}{dt}\Big\vert_{t=0}
\exp tX\circ g\right)
\end{equation}
where $X\in {\cal G}$ and
$m\in T^*_gG$ is a linear map of $T_gG\rightarrow {\bf R}$.
The orbit space $\rho^{-1}(x)/G_x$ is well defined
and called a reduced phase space.
Here, $G_x$ is the stabilizer group
of the point  $ x\in {\cal G}^*$.
The above procedure is called a symplectic reduction.
Furthermore, it can be shown that the reduced
phase space may be naturally identified with
the coadjoint orbit
${\cal O}_x\equiv G\cdot x\subset {\cal G}^*$
 \cite{arno}:
\begin{equation}
\rho^{-1}(x)/G_x\cong G/G_x\cong  G\cdot x.
\end{equation}
It can be shown that the reduction can also be achieved by
Dirac's constraint analysis and more detailed analysis
and explicit examples can be found in Ref. \cite{ohpp}.

Let us consider possible types of $SU(N)$ coadjoint orbits
${\cal O}_{\{n_1,n_2,\cdots,n_l\}}
\equiv SU(N)/SU(n_1)\times\cdots\times SU(n_l)\times U(1)^{l-1}$.
Here we have $\sum_{i=1}^ln_i=N$ and the rank of the subgroup
$H\equiv SU(n_1)\times \cdots\times SU(n_l)\times U(1)^{l-1}$
is equal to $N-1$.
It is well known that there is a natural symplectic structure
on the coadjoint orbits of Lie group \cite{kiri}.
They also have the complex structure inherited from
those of $SL(N, {\bf C})$ and   $P_{\{n_1,n_2,\cdots, n_l\}}$
since ${\cal O}_{\{n_1,n_2,\cdots,n_l\}}
=SL(N, {\bf C})/P_{\{n_1,n_2,\cdots, n_l\}}$,
where  $SL(N, {\bf C})$ is the complexification
of $SU(N)$ and $P_{\{n_1,n_2,\cdots, n_l\}}$ is a parabolic subgroup
of  $SL(N, {\bf C})$ which is the subgroup
of block upper triangular matrices in the
$(n_1+n_2+\cdots +n_l)\times (n_1+n_2+\cdots +n_l)$ block decomposition
of an element  of $SL(N,{\bf C})$.
 Borel subgroup $B_N$ corresponds $P_{\{1,1,\cdots, 1\}}$.
Together with the symplectic structure, they
become K\"ahler manifolds.
Let us assume that the symplectic two form is given
in the local complex coordinate $(\bar z, z)$
by the K\"ahler form
\begin{equation}
\omega=\sum_{i,j}\omega_{ij}d z_i\wedge d\bar z_j\label{sympe}
\end{equation}
where $\omega_{ij}$ can be expressed in terms
of K\"ahler potential $W$ by
\begin{equation}
\omega_{ij}=i\partial_i\bar\partial_j W,
\end{equation}
Then the Poisson bracket can be defined via
\begin{equation}
\{F,G\}=\sum_{i,k}\omega^{ki}\left(\frac{\partial
F}{\partial  z_k}\frac{\partial G}{\partial\bar z_i}-
\frac{\partial G}{\partial z_k}
\frac{\partial F}{\partial\bar z_i}\right)
\end{equation}
where the inverse  $\omega^{ki}$ satisfies
$\omega_{ik}\omega^{kj}=\delta_i^j$.

The plan of the paper is as follows.
In Section 2,  we give a classical
description of integrable spins on
coadjoint orbits. A complete set of commuting functions
on each orbit
and  construction of  a concrete integrable system on the
maximal orbit of $SU(3)$ group is given. In Section 3,
we quantize the system by using
the technique of coherent state and solve explicitly the
time-dependent Schr\"odinger equation which is set up
by geometric quantization method.
Section 4 contains conclusion and discussions.

\def\theequation{\arabic{section}.\arabic{equation}}
\section{Integrable Spin on $SU(3)$ Flag Manifold}
\setcounter{equation}{0}

Let us discuss about integrable models on
${\cal O}_{\{n_1,n_2,\cdots,n_l\}}$.
We denote $F_a$'s  the Hamiltonian functions
associated with the vector fields $T_a$'s
generated by  the generators
$X_a$'s, $[X_a,X_b]=f_{abc}X_c$:
\begin{equation}
T_a\rfloor\omega = dF_a,
\label{vector}
\end{equation}
and these $F_a$'s satisfy the Poisson-Lie
relations \cite{arno}:
\begin{equation}
\{F_a, F_b\}=f_{abc}F_c.
\label{opp}
\end{equation}
The $f_{abc}$'s  are structure constants of the group $G$.
To construct  commuting functions on
${\cal O}_{\{n_1,n_2,\cdots,n_l\}}$,
first note that there exist $N-1$ commuting Hamiltonian
functions $F_3, F_8, F_{15}, \cdots, F_{N^2-1}$ which
is obvious from Eq. (\ref{opp}).
These functions leave the maximal torus
$T^{N-1}\subset {\cal O}_{\{n_1,n_2,\cdots,n_l\}}$ invariant.
Let us denote rank $n$ Casimir invariant of
$su(m)$ algebra by $C_n(m)$.
 For example, $C_2(3)=\frac{1}{2}(F_1^2+
F_2^2+\cdots+F_8^2)$ and $C_2(2)=\frac{1}{2}(F_1^2+
F_2^2+F_3^2)$. Then, using the Eq. (\ref{opp}) again,
it can be inferred \cite{oh1} that the functions
$C_p(q)-C_p(q-1),\ 2\leq p\leq q=2,3,\cdots,N-1$
are the other commuting functions.
Here we define $C_p(q)=0$ for $q<p$. So we have a set of
commuting functions which is  given by
\begin{equation}
F_3, F_8, F_{15}, \cdots, F_{N^2-1},
F_{pq}\equiv C_p(q)-C_p(q-1).
\label{commute}
\end{equation}

Note that in deriving the above set of commuting functions,
we only used the Poisson-Lie relations (\ref{opp}) of
$SU(N)$ symmetry on each orbit ${\cal O}_{\{n_1,n_2,\cdots,n_l\}}$
and the number of commuting functions is $N^2-N$ which equals
to the half of the maximal orbit. This makes us suspect that
the commuting functions (\ref{commute}) are not independent
on an arbitrary orbit. An example can be given in
${\cal O}_{\{N-1,1\}}=SU(N)/SU(N-1)\times U(1)$ which is
 the complex projective space $CP(N-1)$.
The complete set of commuting functions is
given by  $F_3, F_8, F_{15}, \cdots, F_{N^2-1}$
and all the $F_{pq}$'s can be expressed in terms of
$F_3, F_8,\cdots,F_{N^2-1}$. More explicitly,
the Hamiltonian  functions $F_a$'s are given by \cite{koh,kimoh}
\begin{equation}
F_a(z,\bar z)=im\sum_{I,K=0}^{N-1}
{\bar Z}_I(X_a)_{IK}Z_K\label{deff2}
\end{equation}
where $m$ is an integer and $Z_0$ and $Z_i$ are given
 in terms of the complex coordinates
$z_i$ on $CP(N-1)$ as follows:
\begin{equation}
 Z_0=\frac{1}{\sqrt{1+\vert z\vert^2}},\ \ \
Z_i=\frac{z_i}{\sqrt{1+\vert z\vert^2}}.
\label{slp}
\end{equation}
It can be easily checked that in $SU(3)$ case, for instance,
we have
\begin{equation}
F_{22}=C_2(2)=\frac{m^2}{72}(2+\frac{\sqrt{12}}{m}F_8)^2.
\end{equation}
So only $F_3$ and $F_8$ are independent commuting functions.

It seems to be a complicated  matter to discuss about which commuting
set of functions should be chosen.
Instead we give a couple of  examples.
Let us first consider the orbit ${\cal O}_{\{n,1,\cdots,1\}}$.
It seems natural to choose the following
as the complete set of commuting functions:
$F_3, F_8, \cdots, F_{N^2-1}$,
 $F_{pq}$ with the constraint $p\leq N-n$.
 $n=N-1$ corresponds to $CP(N)$ orbit (we define $C_1(q)=0$).
 $n=1$ corresponds to the
the maximal orbit
${\cal O}_{\{1,1,\cdots,1\}}=SU(N)/U(1)^{N-1}$ which is
usually called a flag manifold.
The set of independent commuting functions is given by
the Eq. (\ref{commute}).
For example, in $SU(4)$ case, we have six commuting
functions $F_3, F_8, F_{15}, F_{22}=1/2(F_1^2+F_2^2+F_3^2),
F_{23}=1/4(F_4^2+F_5^2+F_6^2+F_7^2+F_8^2), F_{33}=
d_{abc}F_aF_bF_c$, where $d_{abc}$'s are the symmetric
structure constants of $su(3)$.
This case corresponds to
the so-called non-commutative integrability \cite{fome}
in contrast to the Liouville integrable
system of $CP(N-1)$.
If we consider  $\tilde G=SU(N-1)\times U(1)$
group action on  ${\cal O}_{\{1,1,\cdots,1\}}$ and its algebra
$\tilde{\cal G}$ generated by the Hamiltonian functions,
they satisfy the criteria for
the   non-commutative integrability:
\begin{equation}
\mbox{dim}\ \tilde{\cal G}+\mbox{rank}\
\tilde{\cal G}=\mbox{dim}\ {\cal O}_{\{1,1,\cdots,1\}}.
\end{equation}
The level set $M=\{x\in M: F_i=c_i, \ \ X_i\in
su(N-1)\times u(1)\}$
is a smooth $N-1$ dimensional torus $T^{N-1}$.

In the rest of the paper, we will
give an explicit construction of an integrable model
on the $SU(3)$ flag manifold ${\cal O}_{\{1,1\}}$.
Then we will try to quantize the system by  the use of
coherent state quantization method.
We consider the integrable system with $F_3, F_8$, and
$C_2(2)=1/2(F_1^2+F_2^2+F_3^2)$
 in involution with
Hamiltonian given by
\begin{equation}
H\equiv H(F_3, F_8, C_2(2)).
\label{hami}
\end{equation}
The well-known quantization of the above Hamiltonian
is to pursue a procedure within the framework of geometry
of coadjoint orbits \cite{kiri} and Borel-Weil-Bott theory \cite{helg}
or to calculate the unitary representations of $SU(3)$ group
restricted to Cartan subgroup.
We will pursue the coherent state quantization method \cite{klau}
in this paper.
One of the motivation could be that the coherent state path
integral formulation of the generalized spins \cite{ston,oh,oh1}
exhibit many interesting features
which the aforementioned quantization schemes do not have.
More importantly in this paper, the motivation is
that although the general results, for example,
representation theory of the compact $SU(3)$, are well-known,
we are interested in more concrete examples
in which the time-dependent Schr\"odinger equation can be written down
explicitly and its solutions can be looked into.
We find that the coherent state method
provides a convenient tool for this.
We will restrict our Hamiltonian to be at most a
quadratic function of $F_3$ and $F_8$ in which case the analysis
becomes more transparent.
More general cases could be
handled in the same manner, although it could lead to  a more
complicated situation. Let us consider
\begin{equation}
H=\sum_{m,n=3,8}C^{mn} F_mF_n+
\sum_{m=3,8}D^mF_m+\gamma
\label{qhami}
\end{equation}
where $C^{mn}$'s  and $D^m$'s are constants.
Note that the above system describes a type of
generalized spinning tops.
In order to find an explicit expression for the
Hamiltonian given above, we have
to coordinatize the flag manifold ${\cal O}_{\{1,1\}}$.
The ideal choice
for the explicit  construction of the symplectic structure
seems to be the Bruhat coordinatization \cite{helg}.
 According to Bruhat cell decomposition,
the flag manifold ${\cal O}_{\{1,1\}}$ can be covered with six
coordinate patches. The convenient thing about
Bruhat cell decomposition is that the largest
cell provides a coordinatization $(z_1,z_2,z_3)$
of nearly all of the flag manifold missing only
lower-dimensional
subspaces.

The largest cell on ${\cal O}_{\{1,1\}}=SL(3,{\bf C})/B_3$ is
represented as follows \cite{pick}:
\begin{equation}
[g_c(z)]_{B_3}=\left[\left(\begin{array}{ccc}
1&0&0\\
z_1&1&0\\
z_2&z_3&1\\
\end{array}
\right)\right]_{B_3}\mapsto (z_1,z_2,z_3)
\end{equation}
with $g_c\in SL(3,{\bf C})$.
Symplectic structure is given
by the K\"ahler potential $W$
which was calculated explicitly
 in terms of $z_i$'s as follows \cite{pick}:
\begin{equation}
W=\log (1+\vert z_1\vert^2+\vert z_2\vert^2)^p
(1+\vert z_3\vert^2+\vert z_2-z_1z_3\vert^2)^q
\label{symp}
\end{equation}
where $p,q$ are integers for quantizable orbits.
Using the symplectic structure (\ref{sympe}) expressed in terms of $W$,
 we can calculate the Hamiltonian functions $F_a$'s associated with
 the generators $X_a$'s using Eq. (\ref{vector}) \cite{oh1}.
 The functions  $F_3$ and $F_8$ are given by
\begin{eqnarray}
F_3&=&\frac{p}{2}\frac{2\vert z_1\vert^2+\vert z_2\vert^2}{L_1}+
\frac{q}{2}\frac{\vert z_2-z_1z_3\vert^2-\vert z_3\vert^2}{L_2},
\nonumber\\
F_8&=&\frac{\sqrt{3}p}{2}\frac{\vert z_2\vert^2}{L_1}+
\frac{q\sqrt{3}}{2}\frac{\vert z_2-z_1z_3\vert^2+
\vert z_3\vert^2}{L_2},
\end{eqnarray}
where we defined
\begin{eqnarray}
L_1&=& 1+\vert z_1\vert^2+\vert z_2\vert^2\nonumber\\
L_2&=&1+\vert z_3\vert^2+\vert z_2-z_1z_3\vert^2.
\end{eqnarray}
It is to be mentioned that we ignored some constants in
the above $F_3$ and $F_8$ for convenience.
These will be supplemented when the zero point energies are
discussed in the quantization procedure.
 Let us express the Hamiltonian in terms of
\begin{eqnarray}
Q_1&=&F_3+\frac{1}{\sqrt{3}}F_8=
p\frac{\vert z_1\vert^2+\vert z_2\vert^2}{L_1}+
q\frac{\vert z_2-z_1z_3\vert^2}{L_2}\nonumber\\
Q_2&=&F_3-\frac{1}{\sqrt{3}}F_8=
p\frac{\vert z_1\vert^2}{L_1}-
q\frac{\vert z_3\vert^2}{L_2},
\end{eqnarray}
to achieve the notational simplicity and write
the Hamiltonian (\ref{qhami}) as
\begin{equation}
H=\sum_{m,n=1}^2A^{mn} Q_mQ_n+
\sum_{m=1}^2B^mQ_m+\gamma\label{qqhamilt}
\end{equation}
where $A^{mn}$'s and $B^m$'s are some constants.

\def\theequation{\arabic{section}.\arabic{equation}}
\section{Coherent State Quantization}
\setcounter{equation}{0}

To quantize the above system, we use the coherent state
quantization method \cite{klau}. Let us define
\begin{equation}
\vert z\rangle= \sum_{i=1}^{3}\exp (z_i E_i)\vert 0\rangle
\label{coh}
\end{equation}
where $z=(z_1,z_2,z_3)$, $E_i$'s  are the three positive roots and
 $\vert 0\rangle$ is the highest weight vector corresponding
to the geometry of $SU(3)/U(1)\times U(1) \cite{klau}$.
The normalization for Eq. (\ref{coh}) is chosen so that
\begin{equation}
\langle\bar z^\prime\vert z\rangle=
\exp W(\bar z^\prime,z)=(1+\bar z_1^\prime z_1+
\bar z_2^\prime z_2)^p(1+\bar z_3^\prime z_3+(\bar z_2^\prime -
\bar z_1^\prime \bar z_3^\prime)( z_2- z_1z_3))^q.\label{norm}
\end{equation}
The resolution of unity is expressed as
\begin{equation}
I=\int d\mu(\bar z, z)\exp(-W(\bar z,z))\vert z><\bar z\vert.
\label{unity}
\end{equation}
We note that this definition is different from the usual  one \cite{klau} by
the normalization factor $N=L_1^{-p}L_2^{-q}=\exp(-W)$.
We have chosen this definition here because in the subsequent
analysis, $\bar z$ and $z$
can be treated independently \cite{fadd2} and also it enables one
to choose  the holomorphic (or anti-holomorphic) polarization.

Our main interest lies in the evaluation of the propagator
\begin{equation}
G(\bar z^{\prime\prime},z^\prime;t)
= \langle\bar z^{\prime\prime}\vert e^{-i{\hat H}t}\vert
z^\prime\rangle
\end{equation}
which satisfies the time-dependent
 Schr\"odinger equation
\begin{equation}
i\frac{\partial}{\partial t}G(\bar z^{\prime\prime},z^\prime;t)
=\hat HG(\bar z^{\prime\prime},z^\prime;t)
\end{equation}
in the coherent state quantization.
The explicit differential operator form for the Hamiltonian
which is necessary to  set up the Schr\"odinger equation
can be guessed from the geometric quantization method \cite{wood}.
According to geometric quantization of
classical phase space ${\cal O}_{\{1,1\}}$ with symplectic structure
$\omega$, we quantize classical observable $O\equiv O(F_a)$
in  which   the  functions
$F_a$'s satisfy  the  Poisson-Lie   algebra Eq. (\ref{opp}).
The prequantum operators corresponding to the Hamiltonian
functions $Q_m(\bar z,z)$'s are given by
\begin{equation}
\hat Q_m=-i\nabla_{m}+Q_m
\label{operator}
\end{equation}
where $\nabla_{m}\equiv T_{Q_m}-
iT_{Q_m}\rfloor\theta$ and
$\theta$ is the canonical one form
$\omega=d\theta$ expressed as
\begin{equation}
\theta=i\bar\partial W.\label{theta}
\end{equation}

We will be working in  an anti-holomorphic
polarization in which the differential operator
$\hat Q_m=\hat Q_m(\bar z)$  is given by
\begin{equation}
\hat Q_1(\bar z)=\bar z_1\frac{\partial}{\partial \bar z_1}
+\bar z_2\frac{\partial}{\partial \bar z_2},\
\hat Q_2(\bar z)=\bar z_1\frac{\partial}{\partial \bar z_1}
-\bar z_3\frac{\partial}{\partial \bar z_3}.
\label{zero}
\end{equation}
It can be checked that
\begin{equation}
<\bar z\vert \hat Q_m\vert z>\equiv
<\bar z^\prime\vert \hat Q_m\vert z>
\Big\vert_{ z^\prime\rightarrow z}=
\hat Q_m(\bar z^\prime)<z^\prime\vert z>
\Big\vert_{z^\prime=z}=Q_m(\bar z,z)<\bar z\vert z>
\end{equation}
using the reproducing kernel of the flag manifold ${\cal O}_{\{1,1\}}$
given in Eq. (\ref{norm}).
We note  that $Q_1$ and $Q_2$ generate the following torus action \cite{oh1}
\begin{equation}
Q_1: (z_1,z_2,z_3)\mapsto (e^{i\theta_1}z_1,
e^{i\theta_1}z_2, z_3), \quad
Q_2: (z_1,z_2,z_3)\mapsto (e^{i\theta_2}z_1,z_2,  e^{-i\theta_2}z_3)
\end{equation}
and Eq. (\ref{zero}) should contain  zero point energies. One way of
calculating these would be to try to find a representation of
$\hat F_a$'s which satisfy the full $SU(3)$ algebra:
$[\hat F_a,\hat F_b]=if_{abc} \hat F_c$.
After a straightforward computation using the equation (\ref{operator}),
for example, with $SU(3)$ Gell-Mann structure constants,
we find that the  correct representation
is given by
\begin{equation}
\hat Q_1^\prime(\bar z)=\hat Q_1(\bar z)-(\frac{1}{2}+\frac{1}{3\sqrt{3}})p
-\frac{2}{3\sqrt{3}}q,\ \
\hat Q_2^\prime(\bar z)=\hat Q_2(\bar z)-(\frac{1}{2}-\frac{1}{3\sqrt{3}})p
+\frac{2}{3\sqrt{3}}q.
\end{equation}

Since our Hamiltonian Eq. (\ref{qqhamilt}) is a function of commuting
$\hat Q^\prime_m$'s, there is no normal ordering ambiguity.
We are working in the K\"ahler polarization in which the
Hilbert space is the anti-holomorphic sections of the Hermitian line bundle
\begin{equation}
\nabla_{\partial/\partial z_i}\Psi=0.
\end{equation}
Using  (\ref{theta}), we get $\nabla_{\partial/\partial z_i}\Psi=
\frac{\partial}{\partial z_i}\Psi=0$ and we have an anti-holomorphic section.
Also $G(\bar z^{\prime\prime},z^\prime;t)$ is a
function  of $\bar z^{\prime\prime}$ but not of
$z^{\prime\prime}$.
Substituting the explicit form
of the differential operator $\hat Q_1^\prime$ and $\hat Q_2^\prime$
into the operator version of the Hamiltonian (\ref{qqhamilt}),
we get the following Schr\"odinger equation:
\begin{equation}
i\frac{\partial}{\partial t}G(\bar z^{\prime\prime},z^\prime;t)
= \left(\sum_{i,j=1}^3\alpha_{ij}
\bar z^{\prime\prime}_i\frac{\partial}
{\partial\bar z^{\prime\prime}_i}
(\bar z^{\prime\prime}_j\frac{\partial}
{\partial\bar z^{\prime\prime}_j})
+\sum_{i=1}^3\beta_i\bar z^{\prime\prime}_i\frac
{\partial}{\partial\bar z^{\prime\prime}_i}
       +\gamma^\prime\right)G(\bar z^{\prime\prime},z^\prime;t)
\end{equation}
where $\alpha_{ij}=\sum_{m,n}A^{mn}
a_{mi}a_{nj}, \ \beta_i=\sum_m(2A^{mn}d_n+B^m)a_{mi}$
and $\gamma^\prime=\gamma+A^{mn}d_md_n+B^md_m$. Also,
$a_{ni}$ is a $2\times 3$ matrix whose entries are given by
$a_{1i}=(1,1,0)$, $a_{2i}=(1,0,-1)$
and $d_1=-(\frac{1}{2}+\frac{1}{3\sqrt{3}})p
-\frac{2}{3\sqrt{3}}q$ and $d_2=-(\frac{1}{2}-\frac{1}{3\sqrt{3}})p
+\frac{2}{3\sqrt{3}}q$.

The solution to the above equation can be expressed as follows:
\begin{equation}
G(\bar z^{\prime\prime},z^\prime;t)=
\sum_{n_1,n_2,n_3=0}^{\infty}D_{n_1n_2 n_3}
\prod_{i=1}^3(\bar z^{\prime\prime}_i)^{n_i}
\exp\left\{-i(\sum_{i,j=1}^3
\alpha_{ij}n_in_j+\sum_{i=1}^3
\beta_in_i+\gamma^\prime)t\right\}\label{solution}
\end{equation}
where
\begin{equation}
D_{n_1n_2 n_3}=\prod_{i=1}^3\frac{1}{n_i!}
\left(\frac{\partial}{\partial\bar z^{\prime\prime}_{i}}\right)^{n_i}
G(\bar z^{\prime\prime},z^\prime;0)\Big\vert_{
\bar z^{\prime\prime}_1=
\bar z^{\prime\prime}_2=\bar z^{\prime\prime}_3=0}
\end{equation}
and
\begin{equation}
G(\bar z^{\prime\prime},z^\prime;0)
=(1+\bar z_1^{\prime\prime}z_1^{\prime}
+\bar z_2^{\prime\prime}z_2^\prime )^{p}
(1+\bar z_3^{\prime\prime}z_3^\prime +
(\bar z_2^{\prime\prime}-\bar z_1^{\prime\prime}
\bar z_3^{\prime\prime})
( z_2^\prime-z_1^\prime z_3^\prime))^{q}.
\end{equation}
For $\alpha_{ij}=0$, the Eq. (\ref{solution}) sums into a closed expression
\begin{equation}
G=(1+\bar z_1^{\prime\prime}z_1^{\prime}
e^{i\beta_1t}
+\bar z_2^{\prime\prime}z_2^\prime e^{i\beta_2t})^{p}
(1+\bar z_3^{\prime\prime}z_3^\prime e^{i(\beta_2-\beta_1)t}+
(\bar z_2^{\prime\prime}-\bar z_1^{\prime\prime}
\bar z_3^{\prime\prime})
( z_2^\prime-z_1^\prime z_3^\prime)e^{i\beta_2t})^{q}
e^{-i\gamma^\prime t}.
\end{equation}
We mention that the same result except the zero point energies was
obtained by an explicit evaluation of
the coherent state path integral \cite{oh1}.

Finally, the inner product in the Hilbert space  is defined as
\begin{equation}
<\Psi_1(t)\vert\Psi_2(t)>=\int d\mu(\bar z, z)
\exp(-W(\bar z,z))  \bar\Psi_1(z,t)\Psi_2(\bar z,t)
\end{equation}
using the  Eq. (\ref{unity}).
Then, the wave function in the anti-holomorphic polarization
at arbitrary time is given by
\begin{equation}
\Psi(\bar z^{\prime\prime},t)=\int d\mu(\bar z, z)
\exp(-W(\bar z,z)) G(\bar z^{\prime\prime},z;t) \Psi(\bar z,0)
\end{equation}
It is to be mentioned that the anti-holomorphic polarization was
chosen for convenience. The holomorphic polarization is
equally viable and the analysis can be carried out without much change.

\def\theequation{\arabic{section}.\arabic{equation}}
\section{Conclusion}
\setcounter{equation}{0}
In conclusion, we discussed
about classical spin degrees of freedom on
the coadjoint orbits of $SU(N)$ group and
performed coherent state quantization of the system.
We also considered an explicit example with the case of
the $SU(3)$ flag manifold, set up the Schr\"odinger
equation by the technique of geometric
quantization and found an explicit solution.
As was stressed before, quantization aspects of
the Hamiltonian (\ref{hami}) in terms of
compact coadjoint orbit and representation theory is well-known.
In this paper,
a detailed formulation in the framework of the single component
time-dependent Schr\"odinger equation was provided.
The two approaches are connected by change of basis from
irreducible unitary representation to the coherent
state basis \cite{pere}.
It would be interesting to work out the relation explicitly in $SU(3)$ case.
It would also be interesting to extend the same
quantization procedure to other cases. Possible extensions
would be the one to the coadjoint orbits of other groups
including the non-compact ones and
also to  generalize to a system of many spins
in which the Hamiltonian consists of all the differential
operators $\hat F_a$'s instead of only $\hat F_3$ and
$\hat F_8$.
Finally, generalization to non-relativistic field theory
and their quantization would be another interesting topic to pursue.
These will be reported elsewhere.

\acknowledgments
This work is supported by the KOSEF through
C.T.P. at S.N.U. and Ministry of Education through the
Research Institute of Basic Science (BSRI/95-1419).

\end{document}